\documentclass[aps,showpacs,amsmath,amssymb,twocolumn,nofootinbib]{revtex4-1}

\usepackage{cancel}
\usepackage{cases}
\usepackage{subfigure}
\usepackage{graphicx}% Include figure files
\usepackage{dcolumn}% Align table columns on decimal point
\usepackage{bm}% bold math
\usepackage{color}
\usepackage[colorlinks=true,pdfstartview=FitV,linkcolor=blue,citecolor=blue,urlcolor=blue]{hyperref}
\usepackage[mathlines]{lineno}% Enable numbering of text and display math
\usepackage[dvipsnames]{xcolor}
\usepackage{amsmath}
\usepackage{ytableau}

\usepackage{textcase}

\everymath{\displaystyle}
\begin{document}

\newcommand{\beq}{\begin{equation}}
\newcommand{\eeq}{\end{equation}}
\newcommand{\beqs}{\begin{eqnarray}}
\newcommand{\eeqs}{\end{eqnarray}}

\title{Gravitino Cosmology Helped by a Right Handed (S)Neutrino}

\author{Gongjun Choi,$^{1}$}
\thanks{{\color{blue}gongjun.choi@hotmail.com}}

\author{and Tsutomu T. Yanagida,$^{1,2}$}
\thanks{{\color{blue}tsutomu.tyanagida@sjtu.edu.cn}}

\affiliation{$^{1}$ Tsung-Dao Lee 
Institute, Shanghai Jiao Tong University, Shanghai 200240, China}

\affiliation{$^{2}$ Kavli IPMU (WPI), UTIAS, The University of Tokyo,
5-1-5 Kashiwanoha, Kashiwa, Chiba 277-8583, Japan}

\date{\today}

\begin{abstract}
In this paper, we discuss interesting scenarios resulting from the interplay between the gravitino and the lightest right-handed (s)neutrino. We consider two gravitino mass regimes vastly separated, that is, $m_{3/2}=\mathcal{O}(100){\rm eV}$ and $m_{3/2}\simeq100{\rm GeV}$. For the former case, a significant amount of the entropy production in the cosmological history to dilute the gravitino relic abundance is unavoidable for consistency with the number of satellite galaxies in the Milky way. We will show that the right-handed (s)neutrino can play the role of the heavy particle whose late time decay provides such an additional radiation. For the later case, the gravitino of $m_{3/2}\simeq100{\rm GeV}$ may resolve the $S_{8}$ tension as a decaying dark matter. We will show how the lightest right-handed neutrino and its superpartner can help the gravitino decaying dark matter be equipped with a long enough life time and mass degeneracy with a massive decay product to address the $S_{8}$ tension.
\end{abstract}

\maketitle
\section{Introduction}  
In spite of the null observation of the sparticles (supersymmetric partners) in the LHC to date, the minimal supersymmetric Standard Model (MSSM) is still considered to be the most promising one among various proposals for a new physics beyond the Standard Model (BSM). Its capability to explain intermediate scalar masses in a natural manner as well as to achieve improved unification of gauge couplings of the Standard Model and the radiatively induced electroweak symmetry breaking (EWSB) should be still fully appreciated. 

On the other hand, when the MSSM is extended by $U(1)_{\rm B-L}$ gauge symmetry with three right-handed neutrinos, the model can possibly account for not only the active neutrino masses via the seesaw mechanism~\cite{Yanagida:1979as,GellMann:1980vs,Minkowski:1977sc} but the baryon asymmetry of the universe with the primordial leptogenesis~\cite{Fukugita:1986hr}.\footnote{The importance of $U(1)_{\rm B-L}$ gauge symmetry was clearly explained and stressed in \cite{Wilczeck:1979wc}. }

In this extended MSSM framework, the interplay between gravitino and the lightest right-handed (s)neutrino may give rise to interesting cosmological scenarios. Because two right-handed neutrinos suffice for the success of the seesaw mechanism and the primordial leptogenesis~\cite{Frampton:2002qc}, in principle we may have freedom for parameters which the lightest right-handed neutrino is involved with. Moreover, the fact may strengthen the usefulness of the lightest right-handed neutrino that it is the singlet under the SM gauge group. 

In this work, we discuss two interesting scenarios where the interplay between gravitino and the lightest right-handed (s)neutrino can resolve issues that arise in two observables in cosmology: (i) the number of satellite galaxies $N_{\rm sat}$ in the Milky way and (ii) $
S_{8}$ which characterizes the matter fluctuation amplitude at $8h^{-1}{\rm Mpc}$\footnote{The definition of $S_{8}$  is given by $S_{8}\equiv\sigma_{8}(\Omega_{m}/0.3)^{0.5}$ where $\Omega_{m}$ is the DM and baryon energy density fraction in the total energy density of the universe, and $\sigma_{8}$ is the root mean square of matter fluctuations at $8h^{-1}{\rm Mpc}$}. 

Regarding $N_{\rm sat}$, when the requirement of $N_{\rm sat}\gtrsim63$ for a DM model is applied to a low scale SUSY-breaking scenario with the gravitino mass $m_{3/2}=\mathcal{O}(100){\rm eV}$, we will see that the presence of an early matter dominated (EMD) era is predicted and requires a heavy particle of which decay can produces a significant amount of the entropy release. We will show that the light right-handed (s)neutrino can take the role of the heavy particle with $\sim1{\rm TeV}$ mass and a suppressed Yukawa coupling.

On the other hand, when compared to what results from a $\Lambda$CDM model fit to cosmic microwave background (CMB) data, $S_{8}$ values obtained from weak lensing surveys are smaller by $2-3\sigma$ level~\cite{Heymans:2013fya,Abbott:2017wau,Hikage:2018qbn,Hildebrandt:2018yau} (see, for example, Ref.~\cite{DiValentino:2020vvd} for the current status of the problem and plausible resolutions). Among various proposals to resolve the issue, the late time decaying DM is an interesting possibility. We explore the scenario where $100{\rm GeV}$ gravitino is identified with the decaying DM resolving the $S_{8}$ tension with the help of a massless right-handed neutrino and its superpartner $100{\rm GeV}$ sneutrino.

The outline of this paper is as follows. In Sec.~\ref{sec:model}, we discuss the underlying extended MSSM model. This set-up is to serve as the common framework in which we discuss two cases differentiated by the distinct gravitino masses. In Sec.~\ref{sec:caseI}, we discuss the phenomenological strategy for the gravitino with $m_{3/2}\simeq100{\rm eV}$ to address the missing satellite problem and resulting properties of the associated SUSY model (case I). In Sec.~\ref{sec:caseII}, the parallel discussion for the gravitino with $m_{3/2}\simeq100{\rm GeV}$ and the $S_{8}$ tension is made (case II). Our conclusion is given in Sec.~\ref{sec:conclusion}. In what follows, we would use the notation $N_{i}$ for both a chiral superfield of the right-handed neutrino and its fermion component, $\tilde{N}_{i}$ for the right-handed sneutrino (scalar component of $N_{i}$).

%%%%%%%%%%%%%%%%%%%%%%%%%%%%%%%%%%%

\section{Basic Set-up}
\label{sec:model} 
For both of the case I ($m_{3/2}\simeq100{\rm eV}$) in Sec.~\ref{sec:caseI} and the case II ($m_{3/2}\simeq100{\rm GeV}$) in Sec.~\ref{sec:caseII}, we assume an extension of the MSSM as the underlying particle physics model. For the extension, on top of the MSSM gauge symmetry group, we further assume $U(1)_{\rm B-L}$ gauge symmetry and the discrete $Z_{4}$ symmetry. The latter will be assumed global and local in the case I and the case II respectively. As for the particle content of the model, we extend that of the MSSM by introducing three chiral supermultiplets $N_{i}$s ($i=1,2,3$) for the right-handed neutrinos, two chiral supermultiplets $\Phi$ and $\overline{\Phi}$ of which condensation of scalar components spontaneously breaks $U(1)_{\rm B-L}$ symmetry. These additional fields are taken to be singlets under the MSSM gauge symmetry group. Especially the three right-handed neutrinos are required to cancel the mixed anomalies of $U(1)_{\rm B-L}^{3}$ and $U(1)_{\rm B-L}\times[\rm gravity]^{2}$. We show the quantum number assignment of the particle content of the model under $U(1)_{\rm B-L}$ and $Z_{4}$ in Table.~\ref{table:qn}. For denoting the MSSM matter fields, we borrow the notation of representations of $SU(5)_{\rm GUT}$. Only one of the right-handed neutrinos is assumed charged under $Z_{4}$, which is motivated by phenomenological reasons as discussed in Sec.~\ref{sec:caseI} and
Sec.~\ref{sec:caseII}.
\begin{table}[t]
\centering
\begin{tabular}{|c||c|c|c|c|c|c|c|c|c|} \hline
 & $10$ & $\overline{5}$ & $H_{u}$ & $H_{d}$ & $N_{i=2,3}$ & $N_{1}$ & $\Phi$ & $\overline{\Phi}$ & $X$\\
\hline\hline
$U(1)_{\rm B-L}$ & +1 & -3 & -2 & +2 & +5  & +5  & -10 & +10 & 0\\\hline
$Z_{4}$  & 0 & 0 & 0 & 0 &  0  &  +1 & 0 & 0 & 0\\
\hline
\end{tabular}
\caption{Quantum numbers of chiral superfields for particle contents of the model under $U(1)_{\rm B-L}$ and $Z_{4}$. For denoting the MSSM matter fields, we borrow the notation of representations of $SU(5)_{\rm GUT}$.}
\label{table:qn} 
\end{table}

With the symmetry and particle contents specified above, the superpotential of the model extending that of the MSSM ($W_{\rm MSSM}$) is given by 
\beqs
W&=&W_{\rm MSSM}+\sum_{i=2}^{3}\kappa_{i\alpha}N_{i}L_{\alpha}H_{u}\cr\cr
&+&\sum_{i=2}^{3}\frac{y_{i}}{2}\Phi N_{i}N_{i}+\lambda X(2\Phi\overline{\Phi}-V_{\rm B-L}^{2})\,,
\label{eq:superpotential}
\eeqs
where $L_{\alpha}$ and $H_{u}$ are the chiral superfields for the MSSM lepton and the up-type Higgs $SU(2)_{L}$ doublets ($SU(2)_{L}$ indices are contracted via $\epsilon$-tensor), $\kappa_{i\alpha}$, $y_{i}$, and $\lambda$ are dimensionless coupling constants, $V_{\rm B-L}$ is a $U(1)_{\rm B-L}$ symmetry breaking scale, $\alpha$ is the flavor index ($\alpha=e,\mu,\tau$) and $i$ is the index for the mass eigenstate of the right-handed neutrino. The two right-handed neutrinos $N_{2}$ and $N_{3}$ are taken to be responsible for the active neutrino masses via the seesaw mechanism~\cite{Yanagida:1979as,GellMann:1980vs,Minkowski:1977sc} and the baryon asymmetry via the primordial leptogenesis~\cite{Fukugita:1986hr}. For these two mechanisms to work properly, two right handed-neutrinos suffice~\cite{Frampton:2002qc}.

We end this section by commenting on the mass term for $N_{1}$ and the discrete $Z_{4}$ symmetry. After $U(1)_{\rm B-L}$ breaking, the right-handed neutrino $N_{1}$ can have different masses in forms depending on whether $Z_{4}$ is assumed to be global or gauged. For the former case, a spurion with an appropriate $Z_{4}$ charge can be introduced so that $N_{1}$ may obtain a suppressed mass smaller than those of $N_{2}$ and $N_{3}$. The latter case, however, never allows for the mass term of $N_{1}$ and hence $N_{1}$ remains massless.  In the coming sections, for the phenomenological purpose, we shall assume that $Z_{4}$ is the global symmetry in the case I (Sec.~\ref{sec:caseI}) so that the right-handed neutrino $N_{1}$ can have a suppressed mass compared to $N_{2}$ and $N_{3}$. In contrast, we assume $Z_{4}$ is the gauged symmetry in the case II (Sec.~\ref{sec:caseII}) and thus $N_{1}$ becomes massless.

% ==================================================================

\section{$100$\texorpdfstring{\MakeLowercase{e}V}{eV} Gravitino (Case I)}
\label{sec:caseI} 
\subsection{Phenomenology}
\label{sec:phenomenologyI}
When the SUSY-breaking scale is given such that $m_{3/2}=\mathcal{O}(100){\rm eV}$ holds true, the gravitino becomes the lightest supersymmetric particle (LSP) and starts free-streaming after decoupling from the MSSM thermal bath at the temperature~\cite{Moroi:1993mb}  
\beq
T_{d}\sim{\rm max}\left[m_{\tilde{g}}\,,\,26{\rm GeV}\left(\frac{g_{*}(T_{d})}{100}\right)^{\frac{1}{2}}\left(\frac{m_{3/2}}{1{\rm keV}}\right)^{2}\left(\frac{500{\rm GeV}}{m_{\tilde{g}}}\right)^{2}\right]
\label{eq:Td}
\eeq
where $m_{\tilde{g}}$ is a gluino mass, $g_{*}(T_{d})$ is the effective number of relativistic degrees of freedom at $T_{\rm MSSM}=T_{d}$.

As such, the gravitino serves as WDM today and its the relic abundance in terms of $T_{d}$ is given by
\beq
\Omega_{3/2}h^{2}=\left(\frac{T_{3/2,0}}{T_{\nu,0}}\right)^{3}\left(\frac{m_{3/2}}{94{\rm eV}}\right)=\left(\frac{10.75}{g_{*}(T_{d})}\right)\left(\frac{m_{3/2}}{94{\rm eV}}\right)\,,
\label{eq:omega32}
\eeq
where $\Omega_{3/2}$ is the fraction of the critical energy density contributed by the present gravitino energy density, $h$ is a dimensionless present Hubble expansion rate defined via $H_{0}=100h{\rm km/Mpc/sec}$ and $T_{3/2,0}$ is the present gravitino temperature. The second equality is obtained by considering the MSSM thermal bath entropy conservation between the times of the gravitino decoupling and the active neutrino decoupling in the absence of any entropy production. Given Eq.~(\ref{eq:omega32}) and $g_{*}(T_{d})\simeq230$, it is immediately realized that the gravitino with $m_{3/2}\simeq\mathcal{O}(100){\rm eV}$ makes significant contribution to the current DM energy in the universe with the standard cosmology. 

\begin{figure}[t]
\centering
\hspace*{-5mm}
\includegraphics[width=0.45\textwidth]{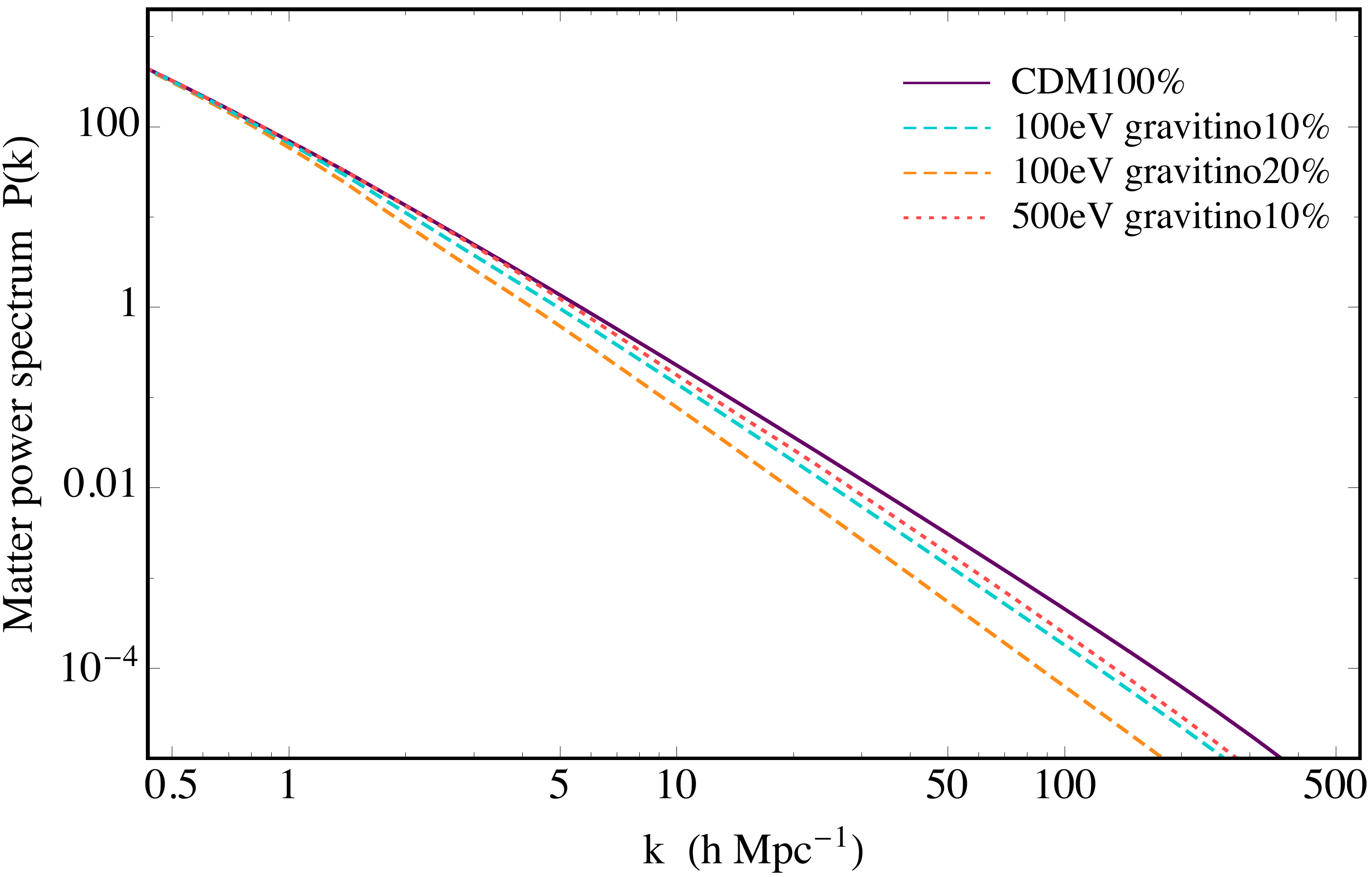}
\caption{The matter power spectrum resulting from the universe distinguished by the different amount and mass of the gravitino WDM. Each $P(k)$ was obtained based on the use of the Boltzmann solver \texttt{CLASS}~\cite{Blas:2011rf} and the choice of different mass and temperature of WDM component in \texttt{ncdm} sector.}
\vspace*{-1.5mm}
\label{fig1}
\end{figure}

Since such a light gravitino mass is expected to produce the free-streaming length of $\mathcal{O}(0.1){\rm Mpc}$, one may wonder if the suppression of matter growth in the small scale ($\lesssim\mathcal{O}(1){\rm Mpc}$) induced by the free-streaming can be consistent with experimental data involved with the small scale structure of the universe. As can be seen in Fig.~\ref{fig1}, the amount of suppression depends on $m_{3/2}$ and $f_{3/2}\equiv\Omega_{3/2}/\Omega_{\rm DM}$ and thus for each $m_{3/2}$, it can be a question if there is a upper bound on $f_{3/2}$ in light of the consistency with the small scale structure of the current universe. 

The data to explore this could include, for instance, the Lyman-$\alpha$ (Ly-$\alpha$) forest observation data, the number of satellite galaxies ($N_{\rm sat}$) in the Milky way, etc. As for the constraint based on the Ly-$\alpha$ data, it was found that about $15\%$ of the total DM relic abundance could be still allowed for the WDM mass greater than $700{\rm eV}$~\cite{Baur:2017stq}.\footnote{For $m_{3/2}\gtrsim700{\rm eV}$, $\Omega_{3/2}h^{2}\gtrsim0.35$ is expected in accordance with Eq.~(\ref{eq:omega32}), which is more than three times greater than $\Omega_{\rm DM}h^{2}\simeq0.12$.} In addition, the authors of Ref.~\cite{Choi:2021lcn} constrained $f_{3/2}$ for $\mathcal{O}(100){\rm eV}$ gravitino WDM based on the requirement $N_{\rm sat}\gtrsim63$ where $63$ is the observed number of satellite galaxies to date~\cite{Polisensky:2010rw,Schneider:2016uqi}\footnote{The requirement $N_{\rm sat}\gtrsim63$ comes from 15 satellite galaxies observed by SDSS (Sloan Digital Sky Survey) with the sky coverage $f_{\rm sky}\simeq0.28$ and 11 classically known satellites. Following Refs.~\cite{Polisensky:2010rw,Giocoli:2007gf,Maccio:2009isa,Horiuchi:2013noa,Kennedy:2013uta,Schneider:2014rda,Schneider:2016uqi,Gariazzo:2017pzb,Diamanti:2017xfo,DEramo:2020gpr}, we take the conservative attitude to take the criteria $N_{\rm sat}\gtrsim63$ although the more recent studies~\cite{Newton:2020cog,Dekker:2021scf} uses $N_{\rm sat}\gtrsim124$ found in \cite{Newton:2017xqg} with the aid of the Dark Energy Survey (DES) data to constrain a DM model. If $N_{\rm sat}\gtrsim124$ is applied, more stringent (larger lower limit) on $\Delta$ would be found. However, even for this case, our model still works since the overlapping of the dashed line corresponding to $\Delta$ as large as $10^{4}$ and the green shaded region was checked. For the current status of the so-called ``missing satellite problem", see \cite{Kim:2021zzw}.} and figured out that the constraint on $\Omega_{3/2}h^{2}$ is smaller than the predicted one given in Eq.~(\ref{eq:omega32}) in the standard cosmology. 

Now given the inconsistency with experimental data reflecting the small scale structure of the universe, the entropy production following an EMD era becomes the inescapable prediction of the low (PeV) scale SUSY-breaking scenario characterized by $m_{3/2}=\mathcal{O}(100){\rm eV}$. With the amount of the entropy production parametrized by $\Delta\equiv s'/s$ where $s'$ ($s$) is the entropy density of the universe after (prior to) a heavy particle decay, it was found that at least $\Delta\simeq10-20$ is required for $N_{\rm sat}\gtrsim63$~\cite{Choi:2021lcn,Diamanti:2017xfo}. Then from the model building point of view, one interesting question could be about a candidate for the heavy particle that can produce the required $\Delta$ via the late time decay. 

The first candidate arising in the minimal scenario could be a messenger particle responsible for the soft SUSY-breaking sparticle mass generation~\cite{Fujii:2002fv,Choi:2020wdq}. Given the relatively low SUSY-breaking scale $M_{\cancel{\rm SUSY}}\simeq\sqrt{m_{3/2}M_{P}}=\sqrt{\mathcal{O}(10^{11}){\rm GeV}^{2}}$ for $m_{3/2}=\mathcal{O}(100){\rm eV}$ with $M_{P}\simeq2.4\times10^{18}{\rm GeV}$ the reduced Planck mass, we expect the soft masses of sparticles to be generated by the gauge mediation~\cite{Dine:1993yw,Dine:1994vc,Dine:1995ag}. Thus one may take the messenger particles as the heavy particle as the source of the entropy production unless the model is extended for the purpose of accommodating a source of the entropy production. 

However, we realize that the messengers cannot be heavy enough to produce the entropy more than $\Delta=10-20$ in $m_{3/2}=\mathcal{O}(100){\rm eV}$ case. Note that the gluino mass generated by the colored messenger-contributed loop needs to meet $m_{\tilde{g}}\gtrsim\mathcal{O}(1){\rm TeV}$ to account for the observed Higgs mass 125GeV~\cite{Yanagida:2012ef}, which puts the upper bound on the messenger mass as $M_{\rm mess}\lesssim\mathcal{O}(10^{6}){\rm GeV}$. In contrast, for the late time decay of the messenger prior to BBN era to lead to $\Delta\gtrsim10-20$, $M_{\rm mess}\gtrsim\mathcal{O}(10^{7}){\rm GeV}$ is required~\cite{Fujii:2002fv}. Therefore, the absence of the range of $M_{\rm mess}$ satisfying these two conditions simultaneously concludes that there should be a source of the entropy production other than the messenger particles.

In the next subsection, as an answer to this question, we will show and suggest that the right-handed neutrino $N_{1}$ shown in Table.~\ref{table:qn} and its superpartner $\tilde{N}_{1}$ can successfully play the role of the heavy particle for the entropy production. Such a possibility is interesting because one may wonder a cosmological role of $N_{1}$ given that two right-handed neutrinos $N_{i=2,3}$ are enough for seesaw mechanism and the primordial leptogenesis as was mentioned in Sec.~\ref{sec:model}.

\subsection{Model I}
\label{sec:modelI}
In this section, we present a mechanism in which the requisite entropy production can occur via the decay of the lightest right-handed neutrino $N_{1}$ and its superpartner sneutrino $\tilde{N}_{1}$. The basic set-up of the model is presented in Sec.~\ref{sec:model}.

As was mentioned in the last paragraph of Sec.~\ref{sec:model}, for the case I, we assume a discrete $Z_{4}$ global symmetry and a spurion $\epsilon$ with $Z_{4}$ charge $-1$. Then aside from the terms given in Eq.~(\ref{eq:superpotential}) we have the following new terms added to the superpotential of the model
\beq
W\supset\kappa_{1\alpha}\epsilon N_{1}L_{\alpha}H_{u}+\frac{y_{1}}{2}\epsilon^{2}\Phi N_{1}N_{1}\,.
\label{eq:Wnew}
\eeq
On the spontaneous breaking of $U(1)_{\rm B-L}$ by the acquisition of a vacuum expectation value (VEV) of $\Phi$ and $\overline{\Phi}$ ($\langle\Phi\rangle=\langle\overline{\Phi}\rangle\equiv V_{\rm B-L}/\sqrt{2}$), $N_{i}$s ($i=1-3$) obtain the masses $m_{1}=(y_{1}\epsilon^{2}V_{\rm B-L})/\sqrt{2}$ and $m_{i=2,3}=(y_{i=2,3}V_{\rm B-L})/\sqrt{2}$. This shows that $m_{1}<\!\!<m_{2},m_{3}$ can be the case for $\epsilon<\!\!<1$. As an exemplary value of $U(1)_{\rm B-L}$ breaking scale leading to a successful explanation of the active neutrino masses via the seesaw mechanism, we take $V_{\rm B-L}\sim10^{15}{\rm GeV}$ close to the GUT scale.

For a reheating temperature satisfying $T_{\rm RH}>m_{2},m_{3}$, the out-of-equilibrium decay of $N_{2}$ and $N_{3}$ (and their superpartners $\tilde{N}_{2}$ and $\tilde{N}_{3}$) produces a primordial lepton asymmetry (thermal leptogenesis). On the other hand, both of $N_{1}$ and $\tilde{N_{1}}$ are produced mainly via the $s$-channel MSSM particle scattering mediated by the gauge boson and the gaugino of $U(1)_{\rm B-L}$. The five dominant relevant scattering events are (1) the scattering among the SM fermions to produce either a pair of $N_{1}$s or a pair of $\tilde{N}_{1}$s ($f_{\rm SM}+f_{\rm SM}^{*}\rightarrow N_{1}(\tilde{N}_{1})+N_{1}(\tilde{N}_{1}^{*})$) (2) the scattering among the MSSM sfermions to produce either a pair of $N_{1}$s or a pair of $\tilde{N}_{1}$s ($\tilde{f}_{\rm SM}+\tilde{f}_{\rm SM}^{*}\rightarrow N_{1}(\tilde{N}_{1})+N_{1}(\tilde{N}_{1}^{*})$) and (3) the scattering between a SM fermion and its superpartner to produce $N_{1}$ and $\tilde{N}_{1}$ ($f_{\rm SM}+\tilde{f}_{\rm SM}^{*}\rightarrow N_{1}+\tilde{N}_{1}$).\footnote{We assume $\langle\tilde{N}_{1}\rangle=0$ during the inflation.} After the gauge boson and the gaugino of $U(1)_{\rm B-L}$ are integrated-out, the scattering rates for these five processes are all given by $\Gamma\propto T^{5}/V_{\rm B-L}^{4}$. Thus, these scattering events are expected to generate equal primordial relic abundances of $N_{1}$ and $\tilde{N}_{1}$. The estimate of the comoving number density of $N_{1}$ generated by the first process is given by~\cite{Kusenko:2010ik}
\beqs
Y_{*}\equiv\frac{n_{1}}{s}&\sim&\left.\frac{n_{\rm SM}\Gamma(f_{\rm SM}+f_{\rm SM}^{*}\rightarrow2N_{1}s)/\mathcal{H}}{s}\right\vert_{T=T_{\rm RH}}\cr\cr
&\sim&(3\times10^{-3})\left(\frac{g_{*}(T_{\rm RH})}{230}\right)^{-\frac{3}{2}}\cr\cr&&\times\left(\frac{V_{\rm B-L}}{10^{15}{\rm GeV}}\right)^{-4}\left(\frac{T_{\rm RH}}{5\times10^{13}{\rm GeV}}\right)^{3}\,,
\label{eq:Y1}
\eeqs
where $n_{1}$ ($n_{\rm SM}$) is the number density of $N_{1}$ (SM fermions), $\mathcal{H}$ is the Hubble expansion rate, and $s$ is the entropy density of the MSSM thermal bath. Because of $s\propto T^{3}$, $n_{\rm SM}\propto T^{3}$ and $\mathcal{H}\propto T^{2}$, it can be easily seen that $N_{1}$ production is most efficient at the reheating era and hence we evaluate $Y_{*}$ at $T=T_{\rm RH}$ in Eq.~(\ref{eq:Y1}). Therefore when taken into account together, the five scattering processes eventually give rise to the following comoving number densities $Y_{1}$ and $\tilde{Y}_{1}$ of $N_{1}$ and $\tilde{N}_{1}$ respectively in terms of $Y_{*}$ 
\beq
Y_{1}\simeq\frac{5}{2}Y_{*}\quad,\quad \tilde{Y}_{1}\simeq\frac{5}{2}Y_{*}\,.
\label{eq:Y2}
\eeq

After the non-thermal production of $N_{1}$ and $\tilde{N}_{1}$, the above comoving number densities in Eq.~(\ref{eq:Y2}) are conserved before the decay of $N_{1}$ and $\tilde{N}_{1}$ take place. And those start to be diluted due to the entropy production when $\Gamma(N_{1}\rightarrow H_{u}+L)$ and $\Gamma(\tilde{N}_{1}\rightarrow L(\tilde{L})+\tilde{H}_{u}(H_{u}))$ become comparable to the Hubble expansion rate. Note that both $N_{1}$ and $\tilde{N}_{1}$ can decay thanks to the first term in Eq.~(\ref{eq:Wnew}). With the small mass difference between $N_{1}$ and $\tilde{N}_{1}$ due to the SUSY-breaking neglected, equating the Hubble expansion rate to the above decay rates gives us the temperature of the thermal bath when $N_{1}$ and $\tilde{N}_{1}$ decays
\beqs
T_{\rm decay}&\simeq&3\left(\frac{\kappa_{1\alpha}\epsilon}{10^{-10}}\right)\left(\frac{m_{1}}{10^{4}{\rm GeV}}\right)^{\frac{1}{2}}{\rm GeV}\cr\cr
&\simeq&0.8\left(\frac{\kappa_{1\alpha}\epsilon}{10^{-10}}\right)\left(\frac{y_{1}\epsilon^{2}}{10^{-12}}\right)^{\frac{1}{2}}\left(\frac{V_{\rm B-L}}{10^{15}{\rm GeV}}\right)^{\frac{1}{2}}{\rm GeV}\,,\nonumber\\
\label{eq:Tdecay}
\eeqs
where $\kappa_{1\alpha}$ denotes the largest one of the three Yukawa coupling constants. 

In the case where the amount of the additionally produced entropy through the decay of $N_{1}$ and $\tilde{N}_{1}$ is so large as to be greater than the entropy of the existing MSSM thermal bath, the relic abundance of the gravitino is subject to the dilution.\footnote{Other ways of the entropy production by a heavy particle can rely on the use of the messenger sector in a GMSB scenario~\cite{Fujii:2002fv} and the SUSY-breaking sector~\cite{Ibe:2010ym}.} We may quantify the amount of the dilution by $\Delta\equiv s'/s$ where $s'$ ($s$) is the entropy density of the universe after (prior to) $N_{1}$ and $\tilde{N}_{1}$ decay. Accordingly, the primordial abundance in Eq.~(\ref{eq:omega32}) is subject to the modification to become
\beq
\Omega_{3/2}^{(\rm after)}h^{2}=\frac{\Omega_{3/2}h^{2}}{\Delta}=\frac{1}{\Delta}\left(\frac{T_{3/2,0}}{T_{\nu,0}}\right)^{3}\left(\frac{m_{3/2}}{94{\rm eV}}\right)\,.
\label{eq:newomega32}
\eeq

\begin{figure*}[htp]
  \centering
  \hspace*{-5mm}
  \subfigure{\includegraphics[scale=0.53]{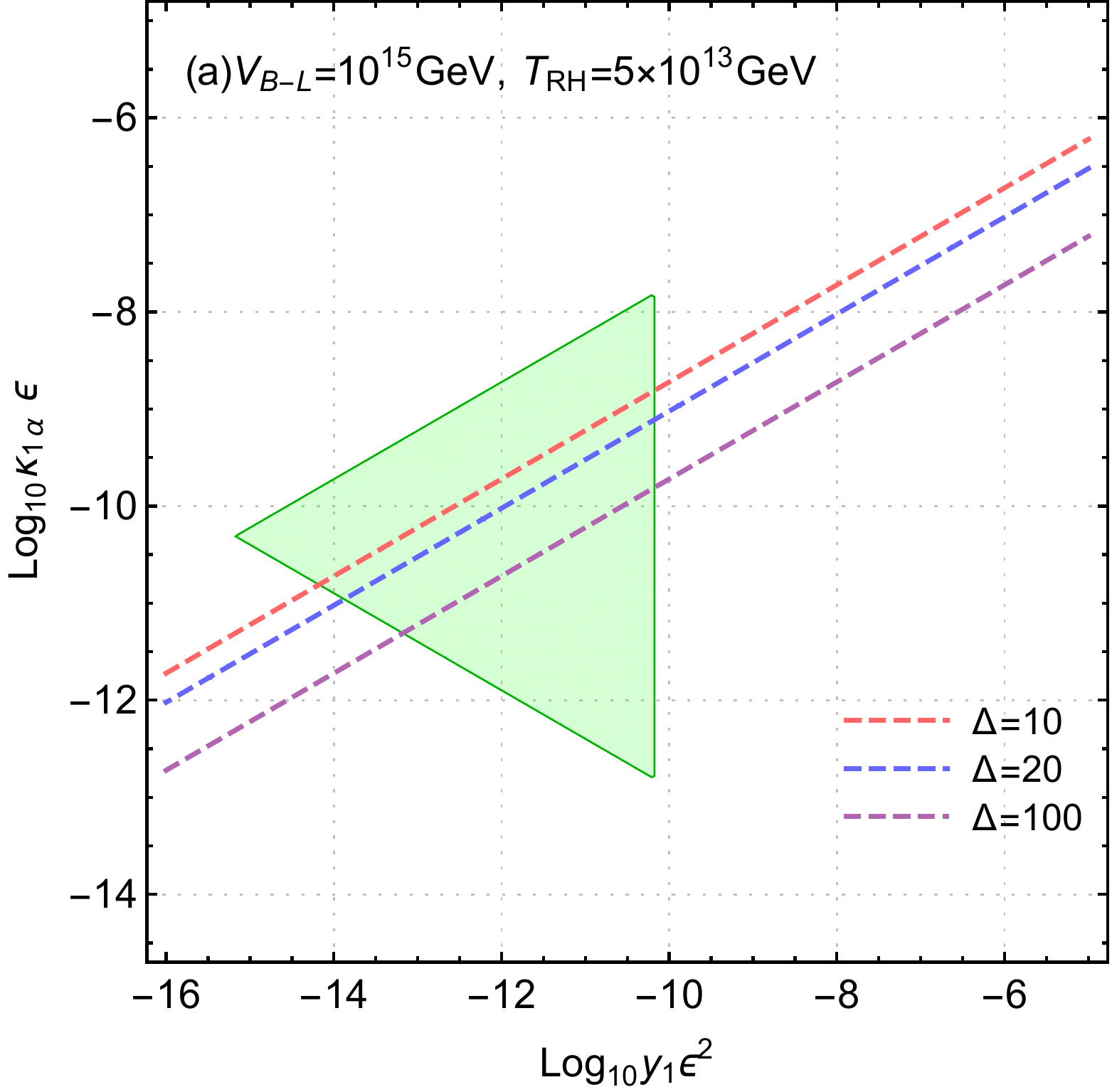}}\qquad
  \subfigure{\includegraphics[scale=0.53]{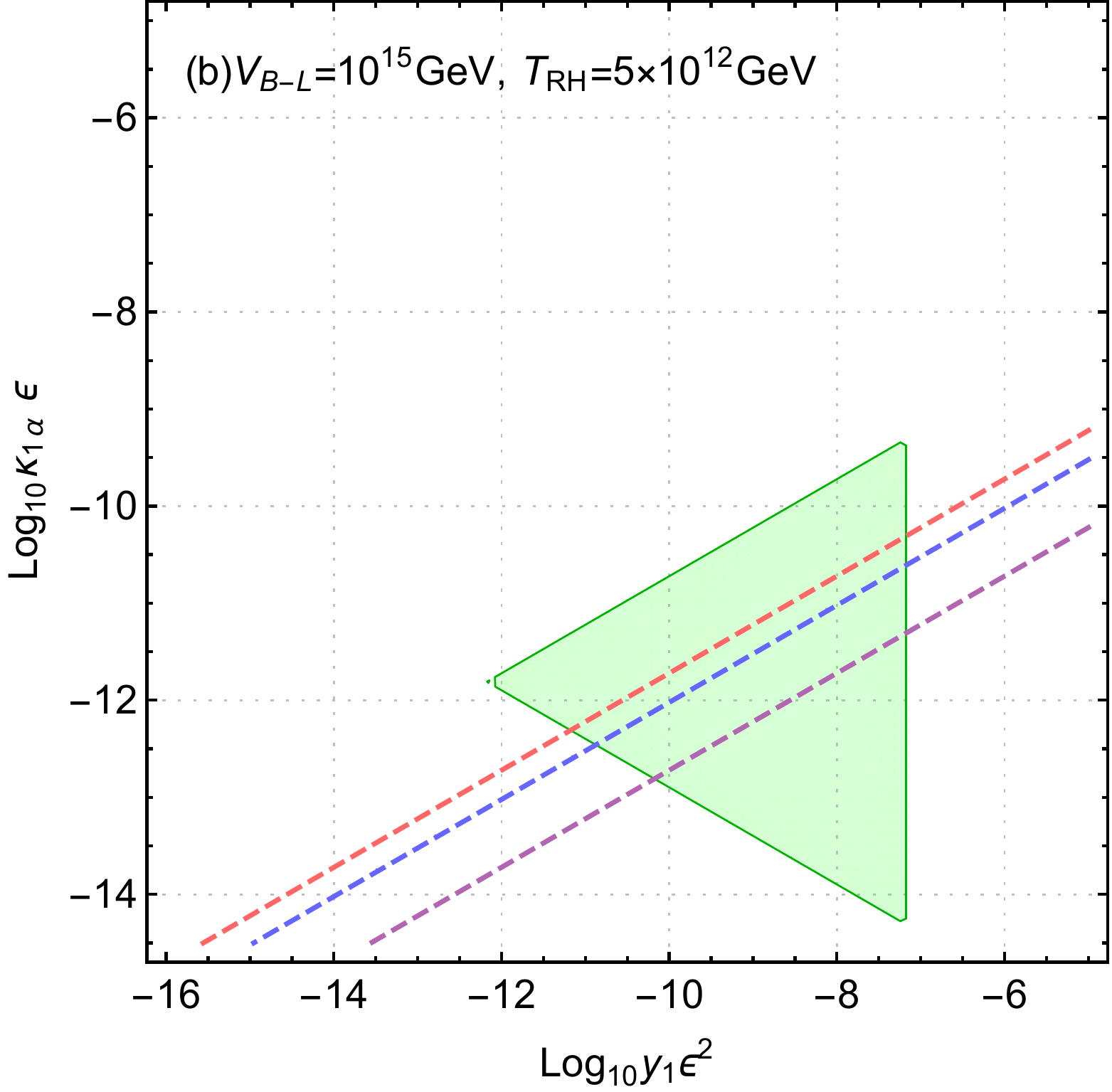}}
  \subfigure{\includegraphics[scale=0.53]{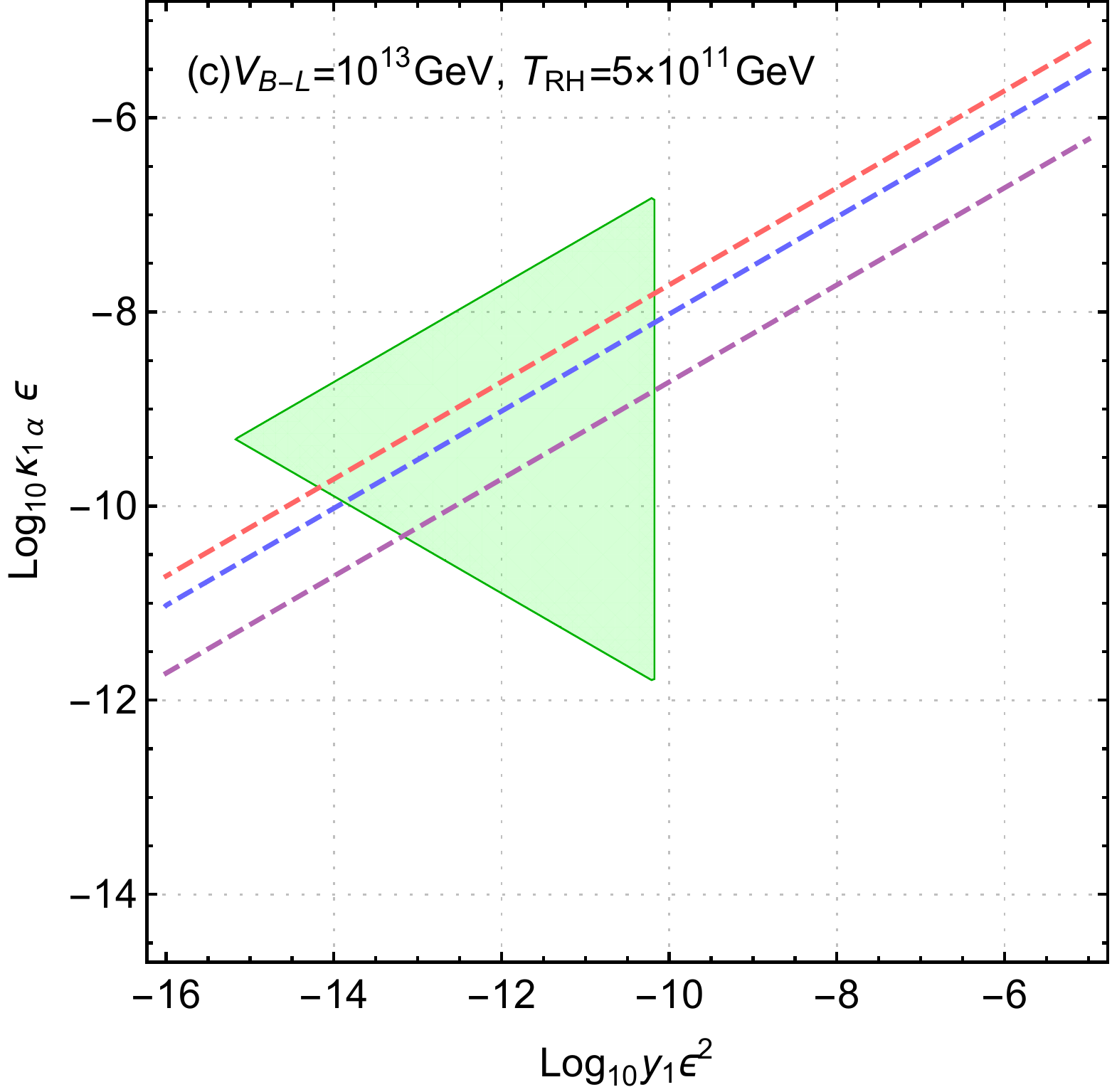}}\qquad
  \subfigure{\includegraphics[scale=0.53]{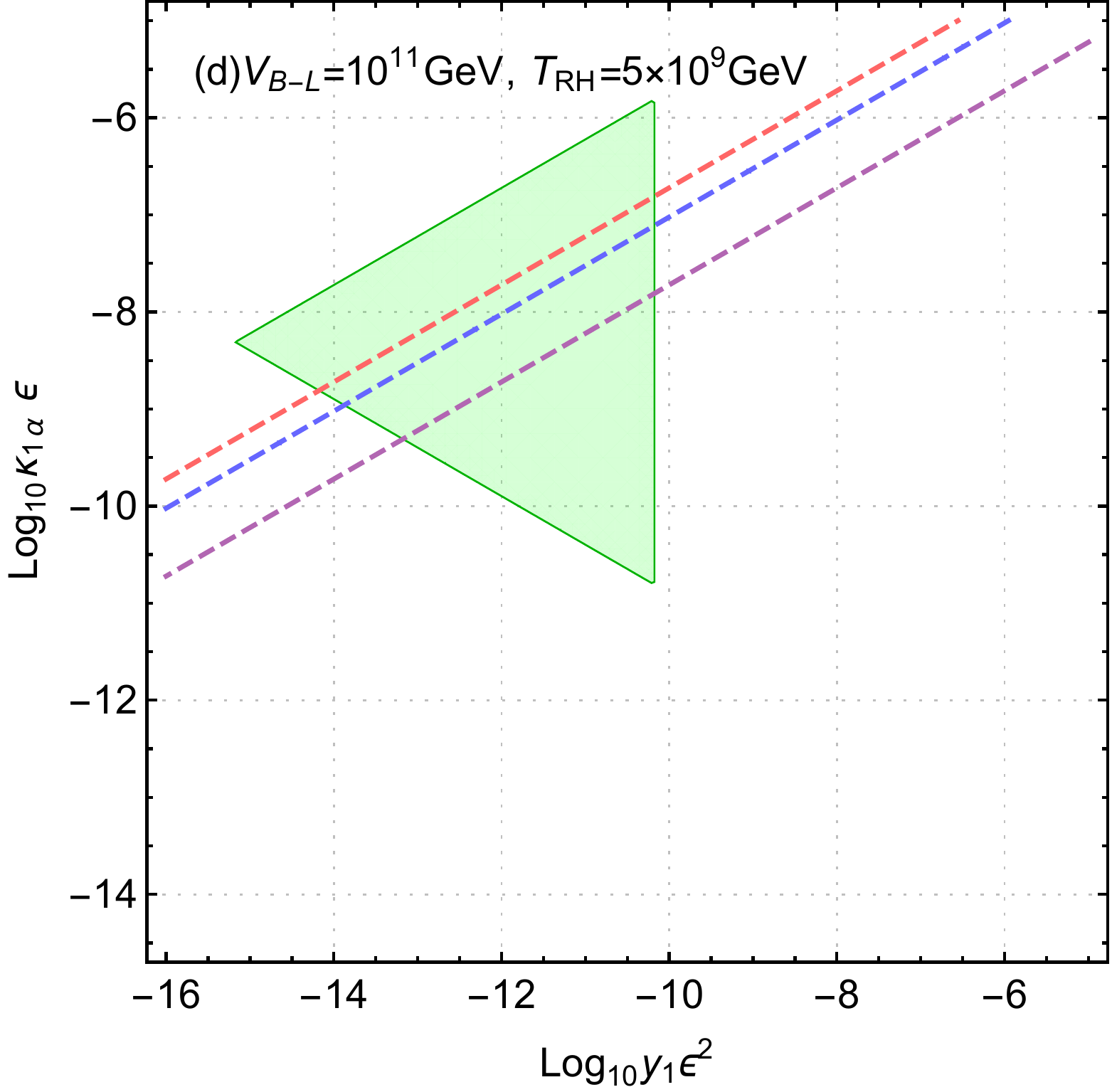}}
  \caption{Parameter space of $(y_{1}\epsilon^{2},\kappa_{1\alpha}\epsilon)$ for a fixed $V_{\rm B-L}$ and a fixed $T_{\rm RH}$. The left and right panels show how the consistent parameter space changes when $T_{\rm RH}$ changes. The green shaded region corresponds to the parameter space which satisfies the condition in Eq.~(\ref{eq:Tc2}). Each of the dashed red, blue and purple is the collection of the points yielding $\Delta=10$, $20$ and $100$ respectively.}
  \vspace*{-1.5mm}
\label{fig:2}
\end{figure*}

This means the temperature ratio decreases by $\Delta^{1/3}$. So in order to have the temperature ratio $\sim0.13-0.15$ today, $\Delta$ is required to be $\sim10-20$.\footnote{As will be discussed, the reheating temperature making our scenario viable could be as high as $T_{\rm RH}\sim5\times10^{13}{\rm GeV}$ for $V_{\rm B-L}=10^{15}{\rm GeV}$. Thus, the baryon asymmetry diluted by the amount of $\Delta\sim10-20$ could be readily compensated by considering a large enough  mass of $N_{2}$ ($\tilde{N}_{2}$) in the thermal leptogenesis thanks to the high $T_{\rm RH}$. See Ref.~\cite{Fujii:2002fv}.}

Now by using $\Delta=s'/s\simeq(4/3)(m_{1}(Y_{1}+\tilde{Y}_{1})/T_{\rm decay})$ and $m_{1}=(y_{1}\epsilon^{2}V_{\rm B-L})/\sqrt{2}$, and referring to Eq.~(\ref{eq:Y1}), Eq.~(\ref{eq:Y2}) and Eq.~(\ref{eq:Tdecay}) we can estimate $\Delta$ as\footnote{Note that in general this expression should apply to the time after a source particle of the entropy production decouples from the primordial MSSM thermal bath.}
\beqs
\Delta&\simeq&19\left(\frac{y_{1}\epsilon^{2}}{10^{-12}}\right)^{\frac{1}{2}}\left(\frac{\kappa_{1\alpha}\epsilon}{10^{-10}}\right)^{-1}\left(\frac{g_{*}(T_{\rm RH})}{230}\right)^{-\frac{3}{2}}\cr\cr&&\times\left(\frac{V_{\rm B-L}}{10^{15}{\rm GeV}}\right)^{-\frac{7}{2}}\left(\frac{T_{\rm RH}}{5\times10^{13}{\rm GeV}}\right)^{3}\,.
\label{eq:Delta}
\eeqs
Observing Eq.~(\ref{eq:Delta}), it is realized that the required $\Delta\sim10-20$ can be readily accomplished indeed for the values of quantities used for the normalization.\footnote{$g_{*}(T_{\rm RH})$ is expected to be greater than $230$ since more degrees of freedom other than the MSSM particles can be found in a SUSY-breaking sector and a messenger sector. Nonetheless, this does not cause any significant change in $\Delta$.} Therefore, we showed that the desired entropy production can be successfully accommodated within the model thanks to the decay of the lightest right-handed neutrino which is necessary for gauging $U(1)_{\rm B-L}$, but not necessary for the successful operation of the seesaw mechanism and the (thermal) leptogenesis. As an example of a parameter set yielding $\Delta=10-20$, we may take ($m_{1},T_{\rm RH},V_{\rm B-L},\kappa_{1\alpha}\epsilon$)=($10^{3}{\rm GeV},5\times10^{13}{\rm GeV},10^{15}{\rm GeV},10^{-10}$) with $g_{*}(T_{\rm RH})=230-300$. 

For consistency of the scenario, we notice that it remains to be checked that the gravitino decouples from the MSSM thermal bath before the summed energy density of the non-relativistic $N_{1}$ and $\tilde{N}_{1}$ dominates the energy budget of the universe, which was the assumption of deriving Eq.~(\ref{eq:Td}). By comparing the energy density of the MSSM thermal bath and that of $N_{1}$ and $\tilde{N}_{1}$, we obtain the following temperature below which the energy of the universe is dominated by that of $N_{1}$ and $\tilde{N}_{1}$
\beqs
T_{c}&=&\frac{4}{3}m_{1}(Y_{1}+\tilde{Y}_{1})\cr\cr
&\simeq&15{\rm GeV}\left(\frac{y_{1}\epsilon^{2}}{10^{-12}}\right)\left(\frac{g_{*}(T_{\rm RH})}{230}\right)^{-\frac{3}{2}}\left(\frac{V_{\rm B-L}}{10^{15}{\rm GeV}}\right)^{-3}\cr\cr
&&\times\left(\frac{T_{\rm RH}}{5\times10^{13}{\rm GeV}}\right)^{3}\,.\nonumber \\
\label{eq:Tc}
\eeqs
Thus for the exemplary parameter values we are targeting, we see that $T_{c}\simeq15{\rm GeV}$ is obtained and thus confirmed to be smaller than $T_{d}\sim m_{\tilde{g}}$ in Eq.~(\ref{eq:Td}).

Now for consistency of the scenario, we finalize the analysis by imposing the following generic condition
\beq
T_{\rm d}>T_{\rm c}>T_{\rm decay}>10{\rm MeV}\,,
\label{eq:Tc2}
\eeq
where the last inequality was introduced to guarantee the successful big bang nucleosynthesis (BBN). Note that the second inequality was originally intended to demand that $T_{c}$ be greater than the MSSM thermal bath when $N_{1}$ and $\tilde{N}_{1}$ decay. But, since $T_{\rm MSSM}$ differs from a temperature of the thermal bath created from the $N_{1}$ and $\tilde{N}_{1}$ decay by an $\mathcal{O}(1)$ factor, it suffices to require $T_{\rm c}>T_{\rm decay}$. In Fig.~\ref{fig:2}, we show the parameter space of $(y_{1}\epsilon^{2},\kappa_{1\alpha}\epsilon)$ satisfying Eq.~(\ref{eq:Tc2}) as the green shaded region. Shown together as the red (blue) dashed line is the set of the points in the parameter space yielding $\Delta=10$ ($\Delta=20$). For $V_{\rm B-L}=10^{15}{\rm GeV}$, the comparison of the panel (a) and (b) shows that $T_{\rm RH}$ as large as $\sim5\times10^{13}{\rm Ge}V$ is preferred from the model building point of view given the different suppression of $y_{1}\epsilon ^{2}$ and $\kappa_{1\alpha}\epsilon$ by different powers of $\epsilon<1$. On the other hand, in comparison of the panels (a), (c) and (d), we see that the consistent parameter space exists for each different $V_{\rm B-L}$ as far as the ratio $V_{\rm B-L}/T_{\rm RH}$ is similar to the case of the panel (a). For the panels (a), (b) and (d), the eventual viable parameter spaces are understood to be points on the red and blue lines lying in the green shaded region, for which $\kappa_{1\alpha}\epsilon>y_{1}\epsilon ^{2}$ satisfied.

We end this section by commenting on the suppression of the couplings in Eq.~(\ref{eq:Wnew}) made by the spurion $\epsilon$. Observing that each $N_{1}$ is accompanied by a single $\epsilon$, one may regard $\epsilon$ as the analog of the normalization factor for the zero mode of the right-handed neutrino field in 5D theory. In a 5D theory compactified on $S^{1}/Z_{2}$ with the extra coordinate $y\in[0,\ell]$, one may consider a model where the MSSM fields, $\Phi$ and $\overline{\Phi}$ fields are localized on the MSSM brane ($y=0$) while the right-handed neutrino fields $N_{i}$s and the gauge supermultiplet of $U(1)_{\rm B-L}$ are bulk fields. Then for the case where the bulk profile of $N_{i}$s is given by $e^{m_{\rm 5D}y}$ with $m_{\rm 5D}$ the bulk mass, one can see that the 5D model can provide us with the 4D effective theory with the similar structure to our model thanks to the suppressed 4D mass and Yukawa coupling constant of $N_{1}$ induced by $\epsilon\sim \sqrt{2m_{\rm 5D}/(Me^{2m_{\rm 5D}\ell}-M)}\simeq\sqrt{2m_{\rm 5D}/M}e^{-m_{\rm 5D}\ell}$ with $M$ 5D fundamental scale defined via $M_{P}^{2}=M^{3}\ell$. This kind of the set-up was discussed in \cite{Kusenko:2010ik} where the seesaw mechanism remains valid and the lightest right-handed neutrino serves as a DM candidate.

% ==================================================================

\section{100\texorpdfstring{G\MakeLowercase{e}V}{GeV} Gravitino (Case II)}
\label{sec:caseII}
\subsection{Phenomenology}
\label{sec:phenomenology2}
For the case where a CDM candidate is able to decay to a massless and a massive decay product, such a decay has been employed to address various cosmological issues including small scale problems for the sub-galactic scale~\cite{Wang:2014ina,Bae:2019vyh}, $S_{8}$ tension~\cite{Enqvist:2015ara,Abellan:2020pmw} and $H_{0}$ tension~\cite{Berezhiani:2015yta,Vattis:2019efj,Choi:2019jck,Choi:2020tqp,Choi:2020udy}.\footnote{See also Refs.~\cite{Chudaykin:2016yfk,Haridasu:2020xaa,Clark:2020miy} for the criticism on the DDM solution to the $H_{0}$ tension.} The framework of this so-called “decaying dark matter (DDM)” is to assume a mother CDM which decays to a massless radiation and a massive daughter WDM. Given this set-up, it is unavoidable for the decay products to obtain a non-vanishing three momentum. The decay process is parametrized by a life time ($\Gamma_{\rm cdm}^{-1}$) of the mother CDM and a fraction $(\xi)$ of the rest mass of the CDM transferred to the massless decay product. As a result, the four momenta of the massless and the massive decay products are given by $p_{\mu}=(\xi m_{{\rm cdm}}, \overrightarrow{p})$ and $p_{\mu}'=((1-\xi) m_{{\rm cdm}},- \overrightarrow{p})$ respectively where $m_{\rm cdm}$ is a mass of the mother CDM. In addition, the dispersion relation of the massive decay product leads to $(m_{\rm wdm}/m_{\rm cdm})=\sqrt{1-2\xi}$ with $m_{\rm wdm}$ a mass of the massive decay product.

Recently, a study invoking the DDM framework to address $S_{8}$-tension was conducted in Refs.~\cite{Abellan:2020pmw,Abellan:2021bpx} through a Monte Carlo Markov Chain (MCMC) against up-to-date CMB, BAO and uncalibrated SNIa data. Interestingly, it was argued that with the inclusion of the prior on $S_{8}$ value, $S_{8}\simeq0.77$ was obtained within the DDM scenario with $(\Gamma_{\rm cdm}^{-1},\xi)=(56{\rm Gyrs},7\times10^{-3})$ (best-fit), resolving the tension.\footnote{This model, however, cannot resolve the Hubble tension~\cite{Abellan:2020pmw}.} A compelling point in the result of the analysis is that  the decrease in $S_{8}\simeq0.77$ as compared to what is inferred from $\Lambda$CDM model is mostly induced by decrease in $\sigma_{8}$ with $\Omega_{m}$ unaffected, which is better in making agreement with the BOSS galaxy clustering constraint on ($\Omega_{m},\sigma_{8}$) where the degeneracy between the two gets broken~\cite{Heymans_2021}. 

The physics behind this resolution is nothing but the suppression of the growth of the matter fluctuation at $k\sim0.1\!-\!1h{\rm Mpc}^{-1}$ caused by the free-streaming of the massive decay product when compared to the $\Lambda$CDM model prediction. Namely, the way for causing the suppression is identical to that of a usual WDM scenario. Yet, the DDM scenario is distinguished from the typical WDM scenario in that the time-dependency is found in both of the amount of the suppression and the cut-off scale in the matter power spectrum~\cite{Abellan:2020pmw}. For that reason the suppression of the linear matter power spectrum can be delayed until the time as late as $z\sim2-3$ for a large enough $\Gamma_{\rm cdm}^{-1}$ and a small enough $\xi$. In that case, the DDM scenario becomes closer to $\Lambda$CDM model in its effect on CMB power spectrum with $\ell\gtrsim\mathcal{O}(10)$ by avoiding to cause a significant early integrated Sachs-Wolfe effect and in its prediction for the growth rate ($f\sigma_{8}\equiv (d\ln\delta_{m}/d\ln a)\sigma_{8}$) for the time $z\gtrsim1$. With these distinctions between the DDM and the WDM scenarios, the potential significant late integrated Sachs-Wolfe effect on the CMB anisotropy power spectrum for low $\ell$ regime ($\ell\sim10$) is expected to be caused by the late time decay of the DDM and it further distinguishes the DDM and the usual WDM scenarios. 

Now inspired by the above phenomenologically compelling resolution to $S_{8}$, a natural question from the particle physics side is whether there exists a well-motivated BSM model accommodating the DDM scenario addressing the $S_{8}$ tension. Obviously one challenging point in answering the question concerns the non-trivial mass spectrum of the three ingredients. We notice that the best-fit value of $\xi\simeq7\times10^{-3}$ is converted to $1\%$ mass difference between the mother CDM and the massive decay product. In the next section, we demonstrate that the model presented in Sec.~\ref{sec:model} can naturally realize the phenomenological DDM scenario alleviating the $S_{8}$ tension when the assumed $Z_{4}$ symmetry is the gauge symmetry.

\subsection{Model II}
\label{sec:model2}
With the basic set-up presented in Sec.~\ref{sec:model} and the discrete symmetry $Z_{4}$ specified as the gauged one,\footnote{$Z_{4}$ in our model does not suffer from any gauge anomaly.} the model II presented in this section does not have the additional superpotential terms given in Eq.~(\ref{eq:Wnew}). Thus, even after the spontaneous breaking of $U(1)_{\rm B-L}$ at the energy scale around $V_{\rm B-L}\sim10^{15}{\rm GeV}$, the right-handed neutrino $N_{1}$ still remains massless. Because of this, sneutrino $\tilde{N}_{1}$ also remains massless until the SUSY-breaking takes place.

Given this set-up, we find that this situation is well suited for realizing the DDM scenario discussed in Sec.~\ref{sec:phenomenology2}.   $\tilde{N}_{1}$ is expected to obtain a mass once the SUSY-breaking takes place and is mediated to $\tilde{N}_{1}$. Now the soft mass of $\tilde{N}_{1}$ can be very close to a mass of gravitino provided it is dominantly generated by the gravity mediation. Remarkably since $\tilde{N}_{1}$ is the SM gauge singlet, even if we are to explain heavy enough mass spectrum for the sparticles in the MSSM consistent with the null observation of SUSY particles in the LHC by relying on a gauge mediation, the soft-mass of $\tilde{N}_{1}$ can be easily dominated by the gravity mediation contribution provided we assume messengers are singlets under $U(1)_{\rm B-L}$. Along this line of reasoning, we see that the coupling between the gravitino, $\tilde{N}_{1}$ and $N_{1}$ in our model can be an excellent candidate of the concrete particle physics example realizing the DDM scenario resolving the $S_{8}$ tension.\footnote{Another possibility is to consider the MSSM extended by an anomalous global $U(1)_{\rm PQ}$ symmetry (anomalous with respect to either $SU(3)_{c}$ or $SU(2)_{L}$). If the axino is the LSP of the model, the coupling between gravitino, axion and axino can be also invoked to make the gravitino DDM candidate. See Refs.~\cite{Choi:2019jck,Hamaguchi:2017ihw}.}

For mapping to our model the phenomenological best-fit values of $(\Gamma_{\rm cdm}^{-1},\xi)=(56{\rm Gyrs},7\times10^{-3})$, we refer to the rate of the gravitino decay to $\tilde{N}_{1}$ and $N_{1}$~\cite{Hamaguchi:2017ihw}
\beqs
\Gamma(\tilde{G}_{\mu}\rightarrow\tilde{N}_{1}+N_{1})&=&\frac{m^{3}_{3/2}}{192\pi M_{P}^{2}}\cr\cr
&\times&\left[1-\left(\frac{m_{1}}{m_{3/2}}\right)\right]^{2}\left[1-\left(\frac{m_{1}}{m_{3/2}}\right)^{2}\right]^{3}\,.\nonumber \\
\label{eq:gravitino_gamma}
\eeqs
where $m_{1}$ is the soft mass of $\tilde{N}_{1}$. The replacement of the mass ratio $m_{1}/m_{3/2}$ with $\sqrt{1-2\xi}$ and the substituting the best-fit value in Eq.~(\ref{eq:gravitino_gamma}) yield $m_{3/2}\simeq216{\rm GeV}$. The SUSY-breaking scale read from this $m_{3/2}$ amounts to
\beq
m_{3/2}=\frac{|F|}{\sqrt{3}M_{P}}\simeq216{\rm GeV}\quad\rightarrow |F|\simeq\mathcal{O}(10^{21}){\rm GeV}^{2}\,.
\label{eq:m32II}
\eeq

For the gravitino with $m_{3/2}\simeq216{\rm GeV}$ to explain the current DM relic density, we notice that the high enough reheating temperature is necessarily required as shown below. When scattering and decay processes involved with strong and EW gauge interactions and top Yukawa coupling are taken into account for gravitino thermal production, the thermal gravitino relic abundance is given by~\cite{Eberl:2020fml,Rychkov:2007uq}
\beq
\Omega_{3/2}h^{2}=0.217\left(\frac{T_{\rm RH}}{10^{7}{\rm GeV}}\right)\left(\frac{100{\rm GeV}}{m_{3/2}}\right)\left(\frac{m_{\tilde{g}}(\mu)}{10{\rm TeV}}\right)^{2}\,,
\label{eq:gravitinodensity}
\eeq
where $m_{\tilde{g}}(\mu)$ is the running gluino mass and the universal gaugino mass relation was used to write $\Omega_{3/2}h^{2}$ in terms of $m_{\tilde{g}}$. For example, for the gluino mass $m_{\tilde{g}}(\mu)\simeq10{\rm TeV}$ and $\Omega_{\rm DM}h^{2}\simeq0.12$, we see from Eq.~(\ref{eq:gravitinodensity}) that $T_{\rm RH}\simeq10^{7}{\rm GeV}$ is required for the gravitino DM with $m_{3/2}\simeq216{\rm GeV}$.\footnote{See also \cite{Bolz:2000fu,Pradler:2006qh} where the thermal gravitino production rate was computed at the leading order in gauge couplings.} This reheating temperature is consistent with non-thermal leptogenesis. Having $T_{\rm RH}\simeq10^{7}{\rm GeV}$ and $V_{\rm B-L}\sim10^{15}{\rm GeV}$ in mind, it can be easily seen that production of both $N_{1}$ and $\tilde{N}_{1}$ is suppressed since the interaction rate of the main production channels (see discussion in Sec.~\ref{sec:caseI}) $\Gamma\sim T^{5}/V_{\rm B-L}^{4}$ never has a chance to exceed the Hubble expansion rate during the radiation-dominated era. Therefore, the gravitino being the dominating DM candidate cannot be spoiled by $\tilde{N}_{1}$.\footnote{Although the early thermal production of $N_{1}$ and $\tilde{N}_{1}$ are highly suppressed, because of their late time production via the gravitino decay after the recombination epoch, the massless Majorana neutrino $N_{1}$ is expected to be present today as the form of the dark radiation. After numerically solving the time evolution equation for the energy density of $N_{1}$, we find that $\Delta N_{\rm eff}(a=1)$ contributed by $N_{1}$ amounts to $\mathcal{O}(1)$. Referring to \cite{Bringmann:2018jpr}, we checked that this value of $\Delta N_{\rm eff}(a=1)$ is still consistent with CMB power spectrum with our decaying gravitino DM characterized by small enough energy conversion from mother DM to $N_{1}$, slow enough decay and late enough transition time (our model corresponds to $\zeta\simeq6\times10^{-3},\kappa<1,a_{t}=\mathcal{O}(0.1)$ in terms of parameters introduced in \cite{Bringmann:2018jpr}).   }  

On the other hand, the gravitino DM with $m_{3/2}\simeq216{\rm GeV}$ is subject to two potentially problematic issues: the model might cause deviation of the light element amount and formation history from the standard BBN scenario and additional unwanted relic abundance of the non-thermal gravitino DM~\cite{Moroi:1993mb}. To avoid these dangers, we focus on the relatively simple case where the next-to-lightest SUSY particle (NLSP) is the gluino ($\tilde{g}$).\footnote{Note that more precisely speaking, of course, $\tilde{g}$ is the next-to-NLSP (NNLSP) because $\tilde{N}_{1}$ and the gravitino are lighter than $\tilde{g}$. Nevertheless, since a very tiny amount of $\tilde{N}_{1}$ is produced at the late universe after recombination, we treat $\tilde{g}$ as the NLSP in our discussion.} To this end, we may consider the gauge mediation model with a pair of messengers $(\Psi,\overline{\Psi})$ transforming as (${\bf 5},{\bf 5}^{*}$) under $SU(5)_{\rm GUT}$ with their mass satisfying $M_{\rm mess}^{2}>\!\!>F$.\footnote{The inequality $M_{\rm mess}^{2}>\!\!>F$ is required for the stability of the SUSY-breaking vacuum to last for a time longer than the age of the universe~\cite{Hisano:2008sy} in a perturbative gauge mediation model.} We assume that the colored $SU(2)_{\rm L}$ singlet messengers ($D,\overline{D}$) are heavier than $SU(2)_{\rm L}$ doublet messengers ($L,\overline{L}$). Then thanks to the annihiilation driven by the strong interaction, the gluino relic comoving number density ($n_{\tilde{g}}/s$) resulting from the freeze-out process is expected to be much smaller than $\mathcal{O}(10^{-12})$ which is the comoving number density of $100{\rm GeV}$ gravitino DM. Note that even for the case where the neutralino is the DM candidate as LSP and coannihilation with the gluino and gluino-gluino bound state formation are taken into account, the comoving number density is still given as $\mathcal{O}(10^{-14})$~\cite{Ellis:2015vaa}. Therefore, the amount of the potential non-thermal gravitinos produced from the gluino decay is negligible.

Regarding the successful BBN, the gluino mass should be constrained from the requirement that its decay takes place before $T_{\rm MSSM}\simeq10{\rm MeV}$ is reached. From the comparison of the decay rate of gluino to the Hubble expansion rate below,
\beqs
&&\Gamma(\tilde{g}\rightarrow g+\tilde{G}_{\mu})\simeq\frac{1}{48\pi}\frac{m_{\tilde{g}}^{5}}{m_{3/2}^{2}M_{P}^{2}}\simeq\frac{T^{2}}{M_{P}}\simeq\mathcal{H}\cr\cr
&\Longrightarrow& m_{\tilde{g}}=10-20{\rm TeV}\quad {\rm for}\quad T_{\rm MSSM}=10{\rm MeV}\,,\nonumber\\
\eeqs
we see that the mass of messengers ($D,\overline{D}$) is required to satisfy $M_{\rm mess}\lesssim10^{15}{\rm GeV}$. Once this condition is met, the potential problematic hadro-dissociation effects caused by the decay products of $\tilde{g}$ could be avoided and the success of the BBN procedure can be guaranteed. In sum, the model can successfully avoid the so-called cosmological gravitino problem as far as we adopt the gauge mediation model with the gluino NLSP to explain SUSY-breaking soft masses and assume $m_{\tilde{g}}\gtrsim10{\rm TeV}$.

Next, with the successful identification of the vertex for explaining the phenomenologically required specific mass spectra, we are still left with the question for the tiny difference between $m_{3/2}$ and $m_{1}$. In other words, we wonder if there is a theoretically plausible way to account for $m_{1}^{2}\simeq(0.98-0.99)\times m_{3/2}^{2}$. To this end, we consider the possibility where there exists a single UV cutoff in the theory and it is not $M_{P}$, but $\Lambda_{\rm UV}\simeq4\pi M_{P}$ (higher cutoff)~\cite{Ibe:2004mp,Ibe:2006fs}. As an example, this could be the case if we assume a hidden interaction under which the SUSY-breaking field $Z$ is strongly coupled ($g\sim4\pi$), but the MSSM fields are weakly-coupled ($h\sim1$) at the energy scale near $M_{P}$. Then in accordance with the \textrm{Näive} dimensional analysis (NDA)~\cite{Luty:1997fk,Cohen:1997rt}, the effective \textrm{Kähler} potential for the energy scale below $\Lambda_{\rm UV}\simeq4\pi M_{P}$ can be written as\footnote{Graviton being assumed to be weakly coupled under the hidden interaction, the NDA enables us to write down the Einstein-Hilbert action as
\beq
S=\frac{\Lambda_{\rm UV}^{4}}{g^{2}}\int d^{4}x\left(\frac{h^{2}\mathcal{R}}{2\Lambda_{\rm UV}^{2}}\right)=M_{P}^{2}\int d^{4}x\left(\frac{\mathcal{R}}{2}\right)\,,
\eeq
where $\mathcal{R}$ is the Ricci scalar. We may recognize this point as the origin of $\Lambda_{\rm UV}\simeq4\pi M_{P}$ in the higher cutoff hypothesis.}
\beqs
&&K_{\rm eff}^{\rm tot}=\cr\cr
&&\frac{\Lambda_{\rm UV}^{2}}{g^{2}}K^{(Z,Y)}_{\rm eff}\left[\left(\frac{gZ}{\Lambda_{\rm UV}}\right),\left(\frac{gZ}{\Lambda_{\rm UV}}\right)^{\dagger},\left(\frac{hY}{\Lambda_{\rm UV}}\right),\left(\frac{hY}{\Lambda_{\rm UV}}\right)^{\dagger}\right]\cr\cr
&+&\frac{\Lambda_{\rm UV}^{2}}{h^{2}}K^{(Y)}_{\rm eff}\left[\left(\frac{hY}{\Lambda_{\rm UV}}\right),\left(\frac{hY}{\Lambda_{\rm UV}}\right)^{\dagger}\right]\,,\nonumber \\
\label{eq:Keff}
\eeqs
where $Y$ collectively denotes chiral superfields of the MSSM sector, and couplings in $K^{(Z,Y)}_{\rm eff}$ and $K^{(Y)}_{\rm eff}$ are order 1.

Now from the Eq.~(\ref{eq:Keff}), it can be seen that for those sfermions whose masses are dominantly generated by the gravity mediation, the universal soft mass up to a leading correction by higher dimensional operators can be given by
\beq
m_{\tilde{f}}^{2}\simeq(1-\frac{1}{g^{2}})\times m_{3/2}^{2}\,, 
\eeq
if an $\mathcal{O}(1)$ coefficient of the operator $\mathcal{O}\sim(Z^{\dagger}ZY^{\dagger}Y)/\Lambda_{\rm UV}^{2}$ is negative. Now that the sneutrino mass $\tilde{N}_{1}$ is dominantly generated by the gravity mediation in our model, its required mass $m_{1}^{2}\simeq(0.98-0.99)\times m_{3/2}^{2}$ could be understood to be stemming from the large UV-cutoff $\Lambda_{\rm UV}\simeq4\pi M_{P}$.

We finalize this section by commenting on a feature of the higher cutoff hypothesis based on Refs.~\cite{Ibe:2004mp,Ibe:2006fs}. As shown in the above, in the model building the higher cutoff hypothesis can be invoked when one needs a justification for suppression of higher dimensional operators including the SUSY-breaking field $Z$. These operators might result in operators multiplied by $m_{3/2}$ or $m_{3/2}^{2}$. Then the higher cutoff hypothesis can naturally explain a dimensionless coefficient of order $(4\pi)^{-1}$ or $(4\pi)^{-2}$. As an example of the application of this hypothesis, for a SUSY-model where soft masses are dominantly generated by the gravity-mediation, the problematic potential flavor changing neutral current (FCNC) could be avoided with the higher cutoff hypothesis via the suppression of the sfermion mixing terms~\cite{Ibe:2004mp,Ibe:2006fs}.

% ==================================================================

% ==================================================================

\section{Conclusion}
\label{sec:conclusion}
{\color{Green}  }

In this paper, we discussed interesting scenarios where the interplay of gravitino and the lightest right-handed (s)neutrino is useful for accounting for experimental observables including $N_{\rm sat}$ and $S_{8}$. To this end, we took as the basic common set-up the MSSM extended by $U(1)_{\rm B-L}$ gauge symmetry, the discrete $Z_{4}$ symmetry and chiral supermultiplets of $\Phi$, $\overline{\Phi}$ and three right-handed neutrinos $N_{i}$ ($i=1-3$) as shown in Table.~\ref{table:qn}. We considered the case where only two $N_{i}$s ($i=2,3$) are responsible for the seesaw mechanism and the leptogenesis. The remaining lightest right-handed neutrino $N_{1}$, as the only field charged under $Z_{4}$, was invoked for different purposes depending on the cosmological problems at hand. This varied uses of $N_{1}$ were achieved by relying on the different nature of $Z_{4}$, i.e. global or gauged.

To have an enough number of satellite galaxies in the Milky way, there should not be too much WDM component, which is realized as the requirement $N_{\rm sat}\gtrsim63$.  When applied to the low SUSY-breaking scenario with $100{\rm eV}$ gravitino LSP, we figured out that it is necessary to have a mechanism to dilute the gravitino relic abundance. We demonstrated that the decay of the lightest right-handed neutrino $N_{1}$ and sneutrino $\tilde{N}_{1}$ can induce the required dilution provided $Z_{4}$ is chosen to be the global symmetry. For the scenario to have a consistent sparticle mass spectrum, a low scale gauge mediation is demanded for the SUSY-breaking mediation to the visible sector. Since the entropy production is the requisite for this possibility, it might be interesting to ask what cosmological (or astrophysical) probes can investigate whether there has ever been any time of an entropy production causing a temporary increase in the evolution of the Hubble expansion rate. If there is, such a probe can be a way to test the $100{\rm eV}$ gravitino scenario we propose in this work. One possibility is to observe the suppression in the inflationary gravitational wave spectrum for the frequency range $\mathcal{O}(10^{-10}){\rm Hz}-\mathcal{O}(10^{-5}){\rm Hz}$~\cite{Choi:2021lcn}. In addition, the early matter-dominated era before BBN in our model may enhance the production of the primordial black holes if there are.\footnote{We thank K. Inomata for useful discussion for this point.}

For the $S_{8}$ tension, we attended to the decaying DM solution with the parameters $(\Gamma_{\rm cdm}^{-1},\xi)=(56{\rm Gyr},7\times10^{-3})$. We find that our model offers the $100{\rm GeV}$ gravitino as the excellent candidate for the DDM if we take $Z_{4}$ as the gauge symmetry and the higher cutoff hypothesis. We pointed out that if the mass of $\tilde{N_{1}}$ can be dominantly generated by the gravity mediation, $\tilde{N_{1}}$ can serve as  the excellent candidate for the massive warm decay product. Combined with the question about a consistent sparticle spectrum, this possibility requires a gauge mediation model incorporating messengers as heavy as $\sim10^{15}{\rm GeV}$ and charged only under the MSSM gauge symmetry group. Notably we proposed the theoretically justified reasoning for the phenomenologically required fine-tuning for the mass difference between the gravitino and $\tilde{N}_{1}$ by relying on the higher cutoff hypothesis.

% ==================================================================

\begin{acknowledgments}
T. T. Y. is supported in part by the China Grant for Talent Scientific Start-Up Project and the JSPS Grant-in-Aid for Scientific Research No. 16H02176, No. 17H02878, and No. 19H05810 and by World Premier International Research Center Initiative (WPI Initiative), MEXT, Japan. 

\end{acknowledgments}

% ================================================================

\bibliography{main}

\end{document}